\documentclass[useAMS,usenatbib]{mn2e}
\def\spose#1{\hbox to 0pt{#1\hss}}
\def\simlt{\mathrel{\spose{\lower 3pt\hbox{$\mathchar"218$}}
     \raise 2.0pt\hbox{$\mathchar"13C$}}}
\def\simgt{\mathrel{\spose{\lower 3pt\hbox{$\mathchar"218$}}
     \raise 2.0pt\hbox{$\mathchar"13E$}}}

\author[Humphrey et al.]{A. Humphrey$^{1,2}$, F. Iwamuro$^{3}$, M. Villar-Mart{\'{\i}}n$^{4}$, L. Binette$^{5}$, E.C. Sung$^{1}$ \\
$^{1}$Korea Astronomy and Space Science Institute, 61-1 Hwaam-dong, Yuseong-gu, Daejeon, 305-348, Republic of Korea\\
$^{2}$Instituto Nacional de Astrof\'{\i}sica, \'Optica y Electr\'onica (INAOE), Aptdo. Postal 51 y 216, 72000 Puebla, Pue., Mexico (ajh@inaoep.mx)\\
$^{3}$Department of Astronomy, Kyoto University, Kitashirakawa, Kyoto 606-8502, Japan\\
$^{4}$Instituto de Astrof\'{\i}sica de Andaluc\'{\i}a (CSIC), Aptdo. 3004, 18080 Granada, Spain\\
$^{5}$Instituto de Astronom\'{\i}a, Universidad Nacional Aut\'onomo de M\'exico, Ap. 70-264, 04510 M\'exico, DF, M\'exico}
\title[]{The extended ionized gas around the z$=$2.44 radio galaxy MRC 0406-244: the nature of the superbubbles and the optical line brightness asymmetries}

\begin{document}

\date{Accepted 14 July 2009. Received 2009 July 13; in original form 2009 June 9.}

\pagerange{\pageref{firstpage}--\pageref{lastpage}} \pubyear{2009}

\maketitle

\label{firstpage}

\begin{abstract}
In this letter, we investigate the nature of the dramatic `super-bubble' emission structures associated with the z=2.44 radio galaxy MRC 0406-244, using rest-frame optical spectroscopy and an archival {\it HST} NICMOS image.  Based on the optical line ratios and the {\it HST} morphology, we conclude that the gas in the superbubbles is photoionized by the obscured active nucleus.  We suggest that this type of structure might be related to the spatially extended HI absorbers that are detected in front of many high-z radio galaxies.  We also suggest that we may be witnessing the destruction of the extended emission line region.  In addition, we investigate the nature of the emission line brightness asymmetry in MRC 0406-244: we conclude that this asymmetry is due to an asymmetry in the mass of ionized gas, confirming the scenario of McCarthy, van Breugel \& Kapahi, and leading us to reject that of Gopal-Krishna \& Wiita.  

\end{abstract}

\begin{keywords}
galaxies: active; galaxies: ISM; galaxies: high-redshift; galaxies: jets; galaxies: evolution
\end{keywords}

\section[]{Introduction}
Powerful radio galaxies at high redshift (z$\ga$2: HzRG hereinafter) continue to play an important role in cosmological investigations.  They provide a means to identify and study the most massive galaxies in the early universe (e.g. R\"ottgering et al. 1994; Seymour et al. 2007), and represent an opportunity to examine the possible symbiosis between the nuclear activity and the host galaxy (e.g. Nesvadba et al. 2006).  

The HzRG MRC 0406-244 (z=2.44: McCarthy et al. 1991) has been the subject of observations in various wavebands.  Radio images obtained at $\sim$5 and $\sim$8 GHz ($\sim$16 and 28 GHz in the rest-frame) show a triple radio source which has a relatively small projected diameter (82 kpc), and which shows a significant rotation measure (Rush et al. 1997; Carilli et al. 1997).  The presence of warm ionized gas, with a spatial extent of at least 66 kpc, has been revealed through spectroscopy and narrow-band imaging (McCarthy et al. 1991; Eales \& Rawlings 1993; Rush et al. 1997; Pentericci et al. 2001; Taniguchi et al. 2001; Iwamuro et al. 2003; Nesvadba et al. 2008).  The spatially integrated fluxes of the UV-optical emission lines are collectively well explained by photoionization by the active nucleus, with roughly solar metallicity, and an internal extinction A$_{v}<0.5$ mag (Humphrey et al. 2008a).  In addition, a spatially extended absorption feature has been detected in the two-dimensional spectrum of Ly$\alpha$ (Pentericci et al. 2001).  The two-dimensional spectrum of the CIV $\lambda$1549 line is also consistent with having been absorbed (Taniguchi et al. 2001).   Using {\it Spitzer} photometrey, Seymour et al. (2007) have estimated the stellar mass of MRC 0406-244 to be $10^{11.36\pm0.13}$.  

{\it Hubble Space Telescope} ({\it HST} hereinafter) images obtained by Rush et al. (1997) and Pentericci et al. (2001) show spatially resolved emission, with a complex figure-of-eight morphology aligned with the axis of the radio emission.  A likely scenario is that this figure-of-eight structure is a pair of `superbubbles' resulting from ambient interstellar media being swept up by a superwind (McCarthy et al. 1999; Taniguchi et al. 2001) or by the radio source (Nesvadba et al. 2008).  If indeed this is the case, then MRC 0406-244 represents an excellent opportunity to study superwinds and feedback related phenomena.  

In this letter, we investigate the nature of the superbubbles, using long-slit spectroscopy of the rest-frame optical emission of MRC 0406-244.  We also revisit the {\it HST} image of the rest-frame optical emission of Pentericci et al. (2001), and examine the nature of the side-to-side line brightness asymmetry in the extended emission line region.  We assume a flat universe with $H_{0}$ = 71 km $s^{-1}$ Mpc$^{-1}$, $\Omega_{\Lambda}$ = 0.73 and $\Omega_{m}$ = 0.27, resulting in a spatial scale of 8.2 kpc per arcsec at the redshift of MRC 0406-244.  

\begin{figure}
\includegraphics{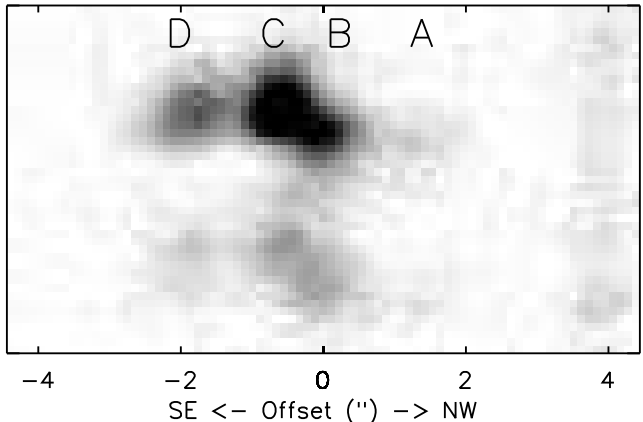} 
\vspace{2.12in}\caption{Section of the two dimensional near-IR spectrum of MRC 0406-244, showing the [OIII] $\lambda\lambda$4959,5007 doublet.  The horizontal axis gives the spatial offset, in arc seconds, relative to the peak of the near-IR continuum emission.  Wavelength increases towards the top of this frame.  The four emission line knots studied in this letter are marked A, B, C and D.} 
\end{figure}

\begin{figure}
\includegraphics{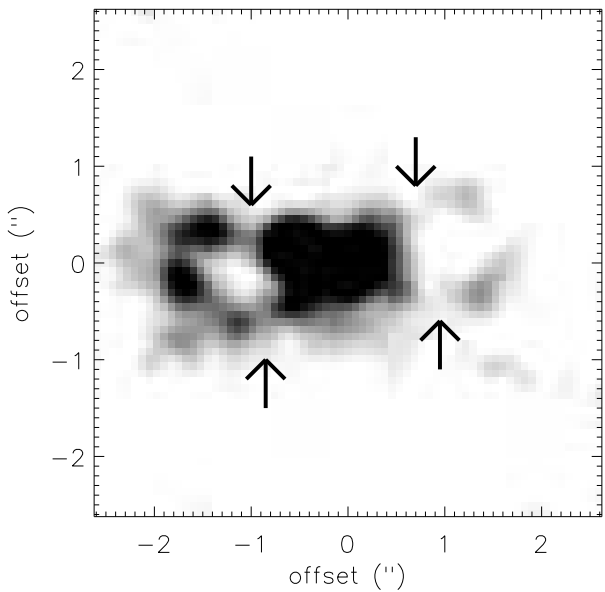} 
\vspace{3.in}\caption{HST NICMOS H-band image of MRC 0406-244, rotated such that the x-axis and the radio axis (PA=128\degr) are parallel to each other.  As discussed in the text, the spatially extended emission is dominated by nebular emission.  The emission minima in the superbubble structures are marked with arrows.  The axes give the spatial offset, in arc seconds, relative to the peak of the H-band emission.} 
\end{figure}

\section[]{Observations and data analysis}

\subsection[]{Subaru near-IR spectrum}
The long slit spectrum of MRC 0406-244 was obtained on 15 September 2000, using the OH Airglow Supressor (OHS hereinafter) spectrograph (Iwamuro et al. 2001) at the Subaru telescope on Hawaii.  The total exposure was 6000 s, with the J and H bands being observed simultaneously.  At the redshift of MRC 0406-244, these two bands sample a rest-frame spectral range of $\sim$3220-3930 \AA~ and $\sim$4290-5240 \AA, respectively.  The slit was 1\arcsec wide, and had a position angle of 132\degr measured North through East, in order to capture the bright emission aligned with the radio source (see e.g. Rush et al. 1997; Pentericci et al. 2001).  The FWHM of the seeing disc was $\sim$0.6\arcsec.  The wavelength scale of the resulting long slit spectrum is 8.5 \AA~ pixel$^{-1}$, and the spatial scale is 0.111\arcsec pixel$^{-1}$.  Iwamuro et al. (2003) give a more in-depth discussion of these observations and their reduction.  The two-dimensional spectrum is shown in Figure 1. 

For the absolute flux calibration, we scaled the spectrum such that the total H-band flux measured in a 2.3\arcsec$\times$1.0\arcsec extraction aperture matches the flux measured in a 2.3\arcsec$\times$1.0\arcsec simulated slit on the H-band image of Iwamuro et al. (2003).  We expect the uncertainty in our absolute flux scale to be $\sim$10 per cent.  We adopt the position of the peak H-band continuum flux as the spatial zero, and we assume that the active nucleus is at this spatial position.  

Four distinct knots of line emission are seen present in the two-dimensional spectrum of MRC 0406-244.  For each of these knots we have extracted a one-dimensional spectrum.  Aperture A (2.16\arcsec NE - 1.05\arcsec NE) captures the knot of emission to the North East of the nucleus; aperture B (0.72\arcsec NE - 0.06\arcsec NE) corresponds to the line emission near the centroid of the continuum emission; aperture C (0.17\arcsec SW - 0.94\arcsec SW) captures the brighter of the two emission line knots on the South West side of the nucleus; aperture D (1.28\arcsec SW - 2.28\arcsec SW) contains the fainter of the two knots to the South West.  We measured fluxes for the emission lines using the {\tt IRAF ONEDSPEC} package.  The 1$\sigma$ uncertainties are dominated by the uncertainty in the level of the continuum.  Our measurements are given in Table 1.  Note that the low spectral resolution of this spectrum ($\sim$1400 km s$^{-1}$) precludes a detailed analysis of the kinematic properties of the emission lines.  

\subsection[]{HST near-IR image}
In addition, we make use of a near-IR {\it Hubble Space Telescope} image of MRC 0406-244 (Proposal ID 7498; PI P. McCarthy), which was taken using Camera 2 of the Near Infrared Camera and Multi-Object Spectrometer (NICMOS hereinafter).  The F160W filter was used, giving a rest-frame wavelength coverage of 4070-5230\AA.  The plate scale is 0.076\arcsec per pixel.  The image is shown in Fig. 2.  For further details, see Pentericci et al. (2001), where this image was first published.

\begin{table*}
\centering
\caption{Emission line fluxes and ratios in the knots along the radio axis of MRC 0406-244.  Columns are as follows: (1) Aperture designation; (2) position in arc seconds of the aperture, relative to the peak of the NIR continuum emission, along the slit; (3)-(7) line fluxes in units of 10$^{-17}$ erg s$^{-1}$ cm$^{-2}$; (8)-(10) emission line ratios referred to in the text.} 
\begin{tabular}{llllllllll}
\hline
Ap. & Pos. & H$\beta$ & [NeV] & [OII] & [NeIII] & [OIII] & [NeIII]/[OIII] & [OIII]/H$\beta$ & [OII]/[OIII] \\
(1) & (2) & (3) & (4) & (5) & (6) & (7) & (8) & (9) & (10) \\
\hline
A & 2.16NE--1.05NE & $\le$3.8     & $\le$4.6    & $\le$12      & $\le$2.3     & 24.6$\pm$1.5  & $\le$0.1 & $\ge$6.1 & $\le$0.5 \\
B & 0.72NE--0.06NE & 5.3$\pm$0.7  & 4.5$\pm$1.5 & 19.0$\pm$2.0 & 6.6$\pm$0.8  & 87.3$\pm$4.4  & 0.076$\pm$0.004 & 17$\pm$2 & 0.22$\pm$0.02 \\
C & 0.17SW--0.94SW & 10.9$\pm$1.1 & 9.7$\pm$1.7 & 27.8$\pm$1.7 & 12.4$\pm$1.1 & 155.2$\pm$7.8 & 0.080$\pm$0.008 & 14$\pm$2 & 0.18$\pm$0.01 \\
D & 1.28SW--2.28SW & 8.1$\pm$1.3  & $\le$3.5    & 29.5$\pm$2.1 & 4.0$\pm$1.9  & 82.2$\pm$4.2  & 0.05$\pm$0.02 & 10$\pm$2 & 0.36$\pm$0.03 \\
\hline
\end{tabular}
\end{table*}

\section[]{Results and discussion}

\subsection[]{Emission line ratios}
Ratios between emission lines are often used to diagnose the physical conditions and the ionization mechanism for the ionized gas associated with active galaxies (e.g. Baldwin, Phillips \& Terlevich 1981).  In all four knots, we find that the optical line ratios are most naturally explained with photoionization by a hard continuum.  In particular, the [NeIII]/[OIII] ratio (Table 1) is consistent with photoionization and A$_{v}$=0 mag (also Humphrey et al. 2008a), but is substantially lower than predicted by fast shock models (i.e., 0.2-1: Dopita \& Sutherland 1996).  The [OIII]/H$\beta$ ratio is also consistent with photoionization by a hard continuum (e.g. Robinson et al. 1987).  The detection of [NeV] in apertures B and C also supports this idea.  This conclusion is in agreement with that of Taniguchi et al. (2001).  It is, however, important to recognise that the line ratios themselves do not allow us to determine the source of the hard ionizing continuum (i.e., the AGN, the metagalactic background radiation or radiative shocks).  

\subsection[]{HST NICMOS image}
Extensive imaging in a variety of wavebands has revealed spectacular filamentary or bubble-like features roughly along the radio axis of MRC 0406-244 (Rush et al. 1997; Pentericci et al. 2001; see Figure 2).  While imaging alone provides insufficient information to elucidate the nature of these morphological features, spectroscopic data can be used to assess what fraction of this spatially extended emission is emitted by warm ionized gas.  For each of our spectroscopic apertures, we first calculated the flux expected due to nebular continuum within the spectral range of the F160W filter.  This we have estimated using the H$\beta$ flux and the nebular continuum emission coefficients in Aller (1984), assuming that the gas has a temperature of 15,000 K.  We then summed the expected nebular continuum flux with the measured flux of line emission, and compared this against the total flux measured within the spectral range corresponding to the wavelength range of the relevant broad-band filter.  In the spatially extended figure-of-eight structure (apertures A, C and D), we find that the overall H-band flux is dominated by nebular emission (i.e. $\ge$90 per cent).  

\begin{figure}
\includegraphics{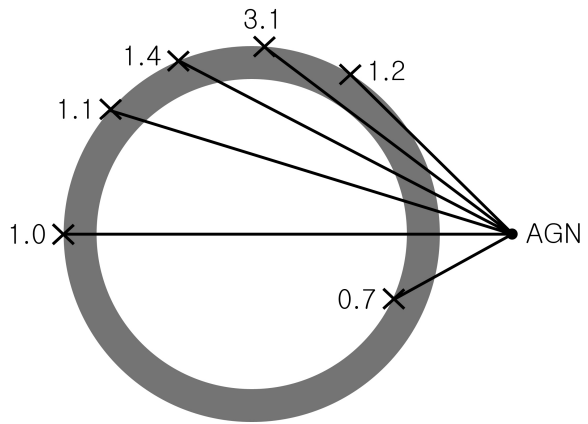} 
\vspace{2.4in}\caption{Diagram to illustrate our simple scenario, in which a uniform spherical shell of gas is irradiated from outside by the active nucleus.  For several points in the shell, we show the relative gas column density measured from that point to the active nucleus (see text).} 
\end{figure}

While Rush et al. (1997) suggested that the figure-of-eight structure has a tidal origin, we prefer the interpretation of McCarthy et al. (1999) and Taniguchi et al. (2001), i.e., this structure is ambient ISM that has been swept up into a bubble by a galactic superwind.  Our reasoning for this is as follows.  In the H-band image, the bright emission knots are arranged into two ellipses, one on each side of the brightest emission peak, which we assume marks the position of the galactic nucleus.  Such morphological features can be naturally explained as gas that has been swept up from the ISM of MRC 0406-244 by a galactic wind, and which has since cooled at the interface between the wind and the external medium.  Indeed, as remarked by McCarthy et al. (1999), these features are reminiscent of the wind-blown bubbles associated with some ultra-luminous infrared galaxies.  In addition, at the centre of each of the emission ellipses there is a region of essentially zero flux.  This can be naturally explained if the wind has been relatively efficient at sweeping up, or at shock-heating, the ambient ISM, leaving little or no cool gas in its wake.  Outside of the bubbles, a significant flux of emission is detected, which can be explained as emission from ISM across which the bubble has not yet expanded.  

In $\S3.1$, we analysed ratios between emission lines, concluding that the gas in the super bubbles is ionized primarily by a hard continuum.  As mentioned in $\S$3.1, the line ratios themselves have not allowed us to discriminate between three potential sources of hard ionizing continuum emission.  In this context, the detailed morphology of the super bubble can provide a useful test to discriminate between photoionization by the anisotrpic radiation field of the AGN, and ionization by shocks or by the isotropic meta-galactic background radiation (MBR).  

Let us suppose that each superbubble in MRC0406-244 is a spherical bubble of cool gas, which is optically thick to ionizing radiation, which has a uniform distribution of material, and which is irradiated by an AGN located at some point just outside of the bubble.  Some points on the bubble will be directly illuminated by the AGN; at other points on the bubble, the incident radiation will have been filtered or absorbed, because the line of sight to the AGN passes through another part of the bubble.  In this simple scheme, the region of the bubble that views the AGN through the highest column density of gas will be at a position angle of $\pm$45-90\degr, measured around the circumference of the bubble from the point closest to the AGN (see Fig. 3).  The nebular emission from this shielded region should be fainter than from other regions of the bubble.  Hence, in the super bubble structures of MRC 0406-244, we would expect to see flux minima in both bubbles at position angles of $\pm$45-90\degr, measured from the point closest to the AGN.  Remarkably, such minima are indeed present in the bubble structures MRC 0406-244 (see Fig 2), and this leads us to suggest that photoionization by the active nucleus is the dominant source of ionization for the EELR of this object.  These minima can also be seen in the {\it HST} F675W and F439W optical images of MRC 0406-244 (Rush et al. 1997).  

The [OI] $\lambda$6300 emission line, which falls into the K-band in the case of MRC 0406-244, would provide a useful test for our scenario.  In the context of emission line regions photoionied by an AGN, the [OI] line is formed in zones that are shielded from ionizing UV photons (these photons having been quenched), but where there remains a strong flux of soft X-ray photons.  In our scenario, these conditions pertain to gas associated with the H-band flux minima in the superbubbles.  Therein, we would expect to measure unusually high [OI]/H$\alpha$ ratios, if indeed our scenario is an accurate representation of `reality'.  Thus, K-band spectroscopy would be useful to test our scenario.  In addition, it would be interesting to image MRC0406-244 in sub-mm continuum and CO line emission, using the next generation of mm-wave interferometers such as ALMA: studying the spatial distribution of the cold ISM should improve our understanding of this galaxy.

\subsection[]{Physical properties}
Having concluded that the bubble structures are photoionized by the active nucleus, it is now possible to obtain an estimate for their gas density $n$ using

\begin{equation}
n = \frac{Q}{4 \pi r^2 c U}
\end{equation}

\noindent where Q is the ionizing photon luminosity of the active nucleus, r is the distance of the gas from the source, and U is the ionization parameter.  Assuming that the ionizing beam of the active nucleus has an opening angle of $\sim$90\degr (Barthel et al. 1989), the global H$\beta$ luminosity of MRC 0406-244 of $L_{H\beta}$=1.6$\times 10^{43}$ erg s$^{-1}$ (Iwamuro et al. 2003; Humphrey et al. 2008a) then implies a minimum Q of 1.3$\times 10^{55}$ s$^{-1}$.  For knot D, at a projected distance of r=12.8 kpc from the AGN, and which has an ionization parameter U=0.004 ($\S$3.1), we thus obtain a gas density of n$\ge$6 cm$^{-3}$.  For knot C, we obtain n$\ge$27 cm$^{-3}$.  These values are consistent with those measured by Nesvadba et al. (2008).  If we assume that the clouds in the bubbles are ionization bounded, then the column density of ionized Hydrogen is 

\begin{equation}
N_{H+} = U c {\alpha_{recB}}^{-1} = 6 \times 10^{20} cm^{-2}
\end{equation}

\noindent where $\alpha_{recB}$ is the total Hydrogen recombination coefficient for case B.  In standard AGN photoionization models $N_{HI}/N_{H} \ga 10^{-6}$, and hence $N_{HI} \ga 6 \times10^{14}$ cm$^{-2}$.  The total mass of ionized Hydrogen in the superbubbles can be estimated using

\begin{equation}
M_{H+} = \frac{L_{H\beta} m_{p}}{h \nu \alpha_{recB} n_{e}}
\end{equation}

\noindent from which we estimate $M_{H+}\sim 6\times 10^{8} M_{\odot}$, assuming that emission from the spatially extended superbubble dominates the total flux of H$\beta$ (see Table 1 col. 3).  It is important to realise that in this calculation we have not considered neutral or molecular gases, and thus the true mass of the superbubbles could be significantly higher.  (See Nesvadba et al. 2008 for a discussion of the energetics of the superbubbles.)

\subsection[]{Line brightness and radio lobe length asymmetries}
There is a trend amongst powerful double radio galaxies in which the spatially extended ionized gas is usually brighter on the side of the nucleus which has the shorter of the two radio lobes (McCarthy, van Breugel \& Kapahi 1991).  The widely accepted explanation for this trend is that the interstellar medium is asymmetrically distributed, resulting in slower expansion of the radio source and brighter line emission on the side of the nucleus where there is more ISM (McCarthy, van Breugel \& Kapahi 1991).  A possible alternative scenario uses motion of the radio galaxy through the inter-galactic medium to produce a lobe-length asymmetry and an asymmetrical extinction of the optical line emission, the latter resulting in a line brightness asymmetry (Gopal-Krishna \& Wiita 1996).  To date, these two scenarios have been subjected to few observational constraints, and very little new data has been applied to understanding the nature of these correlated asymmetries (e.g. Ishwara-Chandra et al. 1998; Humphrey et al 2007).  In this section, we use results from our Subaru long-slit spectroscopy to obtain information about the nature of the line brightness asymmetry in MRC 0406-244.  

As expected, MRC 0406-244 follows the trend, i.e., the optical line emission is brighter on the SE side of the nucleus, which has the shortest radio lobe (e.g. Figs. 1 and 2; Rush et al. 1997).  We investigate the nature of the brightness asymmetry in MRC 0406-244, using two apertures in which the signal to noise of the emission lines is high, which encompass the two brightest emission line knots within the slit, and which are approximately equidistant (in projection) from the assumed position of the active nucleus: apertures B and C (0.72\arcsec NE - 0.06\arcsec NE and 0.17\arcsec SW - 0.94\arcsec SW, respectively; see Table 1).  The main results are as follows.

\noindent 1. We do not detect any significant difference in the redenning sensitive line ratio [NeIII]/[OIII] between B and C, suggesting that there is no significant difference in extinction or reddening by dust between knots B and C.  Therefore, we reject the scenario of Gopal-Krishna \& Wiita (1996), which attempts to explain line brightness asymmetries as an effect of differential extinction by dust.  

\noindent 2. Moreover, several line ratios that are highly sensitive to U, such as [OII]/[OIII] and [OIII]/H$\beta$, show no significant difference between B and C.  This suggests that there is no significant difference between B and C in terms of the local gas density n$_{H}$, or in terms of the local ionizing luminosity Q/4$\pi$r$^{2}$.  (Note: since [OII]/[OIII] is also sensitive to redenning, this result reinforces result 1.)

\noindent 3. Knot C is a factor of $\sim$2 brighter than knot B: taking into account results 1 and 2, the most natural explanation is that there is a higher mass of warm ionized gas on the SE side of the nucleus than on the NW, in agreement with the original scenario suggested by McCarthy, Van Breugel \& Kapahi (1991), and also in agreement with conclusions reached by Humphrey et al. (2007) in the case of the z$=$2.57 radio galaxy TXS 0828+193.  We note that the radio properties of MRC 0406-244 are consistent with this explanation: compared to the NW radio hotspot, the SE hotspot has a lower radio polarization and higher rotation measure (Carilli et al. 1997), consistent with being viewed through a comparatively larger column of magneto-ionized gas.  

Finally, we suggest that high spatial resolution images of emission from molecular gas, neutral Hydrogen and dust would be useful to explore further the nature of the side-to-side asymmetries in this and other powerful radio galaxies.  For instance, does the cold phase of the interstellar medium show a similar brightness asymmetry, and is this correlated with the radio lobe-length asymmetry?

\subsection[]{Implications and speculations}
In MRC 0406-244, the alignment between the superbubbles and the biconical ionizing radiation field of the AGN has allowed the superbubbles to be seen in emission.  It is unclear whether this alignment is by chance or whether the radio/optical axis defines the preferred axis of this kind of outflow.  In this section we ask, what if the superbubbles were not within the ionization cones?  I.e., what if the bubbles expand beyond, or are misaligned with, the cones, and what would be their observational characteristics?  

Being outside of the ionizing radiation field of the active nucleus, the superbubbles would likely have a low ionization fraction, and thus might not be detectable in emission.  However, if the covering factor of the superbubble structure is sufficient (i.e., $\ga$0.1), then the high expected column density of the bubbles, estimated to be $\ga 6 \times10^{14}$ cm$^{-2}$ in the case of MRC 0406-244 ($\S$3.3), would render them detectable via absorption lines such as HI Ly$\alpha$.  We would also expect the absorption lines to be blushifted relative to the line emission from the EELR, due to expansion of the superbubbles, as well as spatially extended.  

Such observation properties are similar to those shown by many of the HI absorbers that are often detected in front of HzRG (e.g. R\"ottgering et al. 1995; van Ojik et al. 1997; Pentericci et al. 2001; Jarvis et al. 2003; Wilman et al. 2004; Villar-Mart{\'{\i}}n et al. 2007; Humphrey et al. 2008b), which leads us to suggest that the superbubbles in MRC 0406-244 and the HI absorbers in front of some other HzRG might be similar phenomena.  

If the bubbles have swept up most of the cool/warm interstellar gas, and if the expansion of the bubbles proceeds at the velocities implied by the emission line kinematics (several hundred km s$^{-1}$: e.g., Taniguchi et al. 2001), then the majority of the extended gas could leave the ionization cones (and host galaxy) on a timescale of $\sim 10^{8}$ yr.  We speculate that we may be witnessing the destruction of the extended emission line region of MRC 0406-244, and the galaxy's transition to join the 1/3 of luminous radio galaxies that emit radio and optical continua, but no detectable emission lines (Miley \& De Breuck 2008).  

\section*{Acknowledgments}
We thank the referee, Patrick McCarthy, for his comments on the manuscript.  LB was supported by the CONACyT grant J-50296.  MVM acknowledges support from the Spanish Ministerio de Educaci\'on y Ciencia through the grants AYA2004-02703 and AYA2007-64712, and from FEDER funds.

\end{document}